\documentclass[twocolumn,preprintnumbers]{revtex4}
\usepackage{amssymb}
\usepackage{amsmath}
\usepackage{graphicx}
\usepackage{dcolumn}
\usepackage{bm}

\begin{document}

\title{Nonlinear organic plasmonics }
\author{B. D. Fainberg\thanks{%
1} }
\affiliation{Faculty of Sciences, Holon Institute of Technology, 52 Golomb St., Holon
58102, Israel }
\affiliation{Raymond and Beverly Sackler Faculty of Exact Sciences, School of Chemistry,
Tel-Aviv University, Tel-Aviv 69978, Israel}
\author{G. Li}
\affiliation{Northwestern University, Department of Chemistry, Evanston IL 60208, USA}

\begin{abstract}
Purely organic materials with negative and near-zero dielectric permittivity
can be easily fabricated. Here we develop a theory of nonlinear
non-steady-state organic plasmonics with strong laser pulses. The
bistability response of the electron-vibrational model of organic materials
in the condensed phase has been demonstrated. Non-steady-state organic
plasmonics enable us to obtain near-zero dielectric permittivity during a
short time. We have proposed to use non-steady-state organic plasmonics for
the enhancement of intersite dipolar energy-transfer interaction in the
quantum dot wire that influences on electron transport through
nanojunctions. Such interactions can compensate Coulomb repulsions for
particular conditions. We propose the exciton control of Coulomb blocking in
the quantum dot wire based on the non-steady-state near-zero dielectric
permittivity of the organic host medium.
\end{abstract}

\maketitle





\section{Introduction}

\label{sec:description}

Metallic inclusions in metamaterials are sources of strong absorption loss.
This hinders many applications of metamaterials and plasmonics and motivates
to search for efficient solutions to the loss problem [1]. Highly doped
semiconductors [1,2]  and doped graphene [3-5] can in principle solve the
loss problem. However, the plasmonic frequency in these materials is an
order of magnitude lower than that in metals making former most useful at
mid-IR and THz regions. In this relation the question arises whether
metal-free metamaterials and plasmonic systems, which do not suffer from
excessive damping loss, can be realized in the visible range? With no
doubts, inexpensive materials with such advanced properties can impact whole
technological fields of nanoplasmonics and metamaterials.

Recently Noginov et al. showed that purely organic materials characterized
by low losses with negative, near-zero, and smaller than unity dielectric
permittivities can be easily fabricated, and propagation of a surface
plasmon polariton at the material/air interface was demonstrated [6]. And
even non-steady-state organic plasmonics with strong laser pulses may be
realized\ [7] that can enable us to obtain near-zero dielectric permittivity
during a short time only.

Approach [6] was explained in simple terms of the Lorentz model for linear
spectra of dielectric permittivities of thin film dyes. However, the
experiments with strong laser pulses [7] challenge theory.

Here we develop a theory of nonlinear non-steady-state organic plasmonics
with strong laser pulse excitation. Our consideration is based on the model
of the interaction of strong (phase modulated) laser pulse with organic
molecules, Ref.[8], extended to the dipole-dipole intermolecular
interactions in the condensed matter. We demonstrate the bistability
response of organic materials in the condensed phase. We also propose the
exciton control of Coulomb blocking [9] in the quantum dot wire based on the
non-steady-state near-zero dielectric permittivity of the organic host
medium using chirped laser pulses.


\section{Model and basic equations}

\label{sec:model}

In this section we shall extend our picture of \textquotedblright
moving\textquotedblright\ potentials of Ref.[8] to a condensed matter. In
this picture we considered a molecule with two electronic states $n=1$
(ground) and $2$ (excited) in a solvent. The molecule is affected by a
(phase modulated) pulse $\mathbf{E}(t)$

\begin{equation}
\mathbf{E}(t)=\frac{1}{2}\mathbf{e}\mathcal{E}(t)\exp (-i\omega t+i\varphi
(t))+\mathrm{c.c.}\text{ \ \ }  \label{eq:E_i(t)}
\end{equation}%
the frequency of which is close to that of the transition $1\rightarrow 2$.
Here $\mathcal{E}(t)$ and $\varphi (t)$ describe the change of the pulse
amplitude and phase in time, $\mathbf{e}$ is unit polarization vectors, and
the instantaneous pulse frequency is $\omega (t)=\omega -d\varphi (t)/dt$.

One can describe the influence of the vibrational subsystems of a molecule
and a solvent on the electronic transition within the range of definite
vibronic transition related to a high frequency optically active (OA)
vibration as a modulation of this transition by low frequency (LF) OA
vibrations $\{\omega _{s}\}$ [10-13]. Let us denote the disturbance of
nuclear motion under electronic transition as $\alpha $. Electronic
transition relaxation stimulated by LFOA vibrations is described by the
correlation function $K(t)=\langle \alpha (0)\alpha (t)\rangle $ of the
corresponding vibrational disturbance with characteristic attenuation time $%
\tau _{s}$ [14-23]. The analytic solution of the problem under consideration
has been obtained due to the presence of a small parameter. For broad
vibronic spectra satisfying the \textquotedblright slow
modulation\textquotedblright\ limit, we have $\sigma _{2s}\tau _{s}^{2}\gg 1$
where $\sigma _{2s}=K(0)$ is the LFOA vibration contribution to a second
central moment of an absorption spectrum. According to Refs. [22,23], the
following times are characteristic for the time evolution of the system
under consideration: $\sigma _{2s}^{-1/2}<T^{\prime }<<\tau _{s}$, where $%
\sigma _{2s}^{-1/2}$ and $T^{\prime }=(\tau _{s}/\sigma _{2s})^{1/3}$ are
the times of reversible and irreversible dephasing of the electronic
transition, respectively. The characteristic frequency range of changing the
optical transition probability can be evaluated as the inverse $T^{\prime }$%
, i.e. $(T^{\prime })^{-1}.$ Thus, one can consider $T^{\prime }$ as a time
of the optical electronic transition. Therefore, the inequality $\tau
_{s}\gg T^{\prime }$ implies that the optical transition is instantaneous.
Thus,\ the condition $T^{\prime }/\tau _{s}<<1$ plays the role of a small
parameter. This made it possible to describe vibrationally non-equilibrium
populations in electronic states $1$ and $2$ by balance operator equations
for the intense pulse excitation (pulse duration $t_{p}>T^{\prime }$). If
the correlation function is exponential: $K(t)/K(0)\equiv S(t)=\exp
(-|t|/\tau _{s})$, the balance operator equations transform into diffusional
equations. Such a procedure has enabled us to solve the problem for strong
pulses even with phase modulation [8,24,25].

Equations of Ref. [8] describing vibrationally non-equilibrium populations
in electronic states $j=1,2$ for the intense chirped pulse excitation,
extended to the dipole-dipole intermolecular interactions in the condensed
matter (see Appendix), take the following form
\begin{eqnarray}
\frac{\partial \rho _{jj}\left( \alpha ,t\right) }{\partial t} &=&\frac{%
\left( -1\right) ^{j}\pi }{2\hbar ^{2}}\delta \lbrack \omega _{21}-p\Delta
n-\omega \left( t\right) -\alpha ]\left( \frac{\varepsilon _{b}+2}{3}\right)
^{2}  \notag \\
&&\times |\mathbf{D}_{21}\mathcal{\vec{E}}(t)|^{2}\Delta ^{\prime }\left(
\alpha ,t\right) +L_{jj}\rho _{jj}\left( \alpha ,t\right)
\label{eq:rhojjfin}
\end{eqnarray}%
where $\Delta ^{\prime }\left( \alpha ,t\right) =\rho _{11}\left( \alpha
,t\right) -\rho _{22}\left( \alpha ,t\right) $, $\Delta n=n_{1}-n_{2}$, $%
\mathbf{D}_{21}$ is the electronic matrix element of the dipole moment
operator. Here $\rho _{jj}$ are the diagonal elements of the density matrix;
$\omega _{21}$ is the frequency of Franck-Condon transition $1\rightarrow 2$%
, and the operator $L_{jj}$ describes the diffusion with respect to the
coordinate $\alpha $ in the corresponding effective parabolic potential $%
U_{j}\left( \alpha \right) $
\begin{equation}
L_{jj}=\frac{1+\left( \alpha -\delta _{j2}\omega _{st}\right) \frac{\partial
}{\partial \left( \alpha -\delta _{j2}\omega _{st}\right) }+\sigma _{2s}%
\frac{\partial ^{2}}{\partial \left( \alpha -\delta _{j2}\omega _{st}\right)
^{2}}}{\tau _{s}}  \label{eq:Ljj}
\end{equation}%
$\delta _{ij}$ is the Kronecker delta, $\omega _{st}$ is the Stokes shift of
the equilibrium absorption and luminescence spectra. The partial density
matrix of the system $\rho _{jj}\left( \alpha ,t\right) $ describes the
system distribution in states $1$ and $2$ with a given value of $\alpha $ at
time $t$. The complete density matrix averaged over the stochastic process
which modulates the system energy levels, is obtained by integration of $%
\rho _{jj}\left( \alpha ,t\right) $ over $\alpha $, $\langle \rho \rangle
_{jj}\left( t\right) =\int \rho _{jj}\left( \alpha ,t\right) d\alpha $,
where quantities $\langle \rho \rangle _{jj}\left( t\right) $ are nothing
more nor less than the normalized populations of the corresponding
electronic states: $\langle \rho \rangle _{jj}\left( t\right) \equiv n_{j}$,
$n_{1}+n_{2}=1$. Furthermore, here $\varepsilon _{b}$ is the
\textquotedblleft bulk\textquotedblright\ relative permittivity (which can
be due distant high-frequency resonances of the same absorbing molecules or
a host medium), $p=\dfrac{4\pi }{3\hbar }|D_{12}|^{2}N$ is the strength of
the near dipole-dipole interaction [26], $N$ is the density of molecules.

Knowing $\rho _{jj}\left( \alpha ,t\right) $, one can calculate the
susceptibility $\chi (\Omega ,t)$ [8] that enables us to obtain the
dielectric function $\varepsilon $ due to relation $\varepsilon (\Omega
,t)=1+4\pi \chi (\Omega ,t)$:%
\begin{align}
\varepsilon (\Omega ,t)& =1+ip_{12}\left( \frac{\varepsilon _{b}+2}{3}%
\right) \{\sqrt{\frac{\pi }{2\sigma _{2s}}}w[\frac{\Omega -(\omega
_{21}-p\Delta n(t))}{\sqrt{2\sigma _{2s}}}]  \notag \\
& -\pi \sigma _{a}\left( \frac{\varepsilon _{b}+2}{3}\right)
^{2}\int_{0}^{t}dt^{\prime }\Delta ^{\prime }\left( \omega _{21}-p\Delta
n\left( t^{\prime }\right) -\omega \left( t^{\prime }\right) ,t^{\prime
}\right)   \notag \\
& \times \tilde{J}(t^{\prime })\sum_{j=1}^{2}\sqrt{\frac{\sigma _{2s}}{%
\sigma \left( t-t^{\prime }\right) }}w[\frac{\Omega +p\Delta n(t)-\omega
_{j}\left( t,t^{\prime }\right) }{\sqrt{2\sigma \left( t-t^{\prime }\right) }%
}]\}  \label{eq:epsilon(Omega)}
\end{align}%
where $\tilde{J}\left( t\right) $ is the power density of the exciting
radiation, $\sigma \left( t-t^{\prime }\right) =\sigma _{2s}\left[
1-S^{2}\left( t-t^{\prime }\right) \right] $, $\omega _{j}\left( t,t^{\prime
}\right) =\omega _{21}-\delta _{j2}\omega _{st}+[\omega \left( t^{\prime
}\right) -\omega _{21}+p\Delta n\left( t^{\prime }\right) +\delta
_{j2}\omega _{st}]S\left( t-t^{\prime }\right) $ are the first moments of
the transient absorption ($j=1$) and the emission ($j=2$) spectra, $\omega
_{st}=\hbar \sigma _{2s}/\left( k_{B}T\right) $ is the Stokes shift of the
equilibrium absorption and luminescence spectra, and%
\begin{equation*}
w(z)=\exp (-z^{2})[1+i\frac{2}{\sqrt{\pi }}\int_{0}^{z}\exp (y^{2})dy]
\end{equation*}%
is the probability integral of a complex argument [27]. It is worthy to note
that magnitude $\varepsilon (\Omega ,t)$ does make sense, since it changes
in time slowly with respect to dephasing. In other words, $\varepsilon
(\Omega ,t)$ changes in time slowly with respect to the reciprocal
characteristic frequency domain of changing $\varepsilon (\Omega )$.

\subsection{Fast vibrational relaxation}

Let us consider the particular case of fast vibrational relaxation when one
can put the correlation function $S\left( t-t^{\prime }\right) $ equal to
zero. Physically it means that the equilibrium distributions into the
electronic states have had time to be set during changing the pulse
parameters. Using Eq.(\ref{eq:rhojjfin}), one can obtain the equations for
the populations of electronic states $n_{1,2}$ in the case under
consideration, which represents extending Eq.(25) of Ref.[8] to the
interacting medium

\begin{eqnarray}
\frac{dn_{1,2}}{dt} &=&\pm \sigma _{a}\left( \frac{\varepsilon _{b}+2}{3}%
\right) ^{2}\exp \{-\frac{\left[ \omega _{21}-p\Delta n-\omega \left(
t\right) -\omega _{st}\right] }{2\sigma _{2s}}^{2}\}  \notag \\
&&\times \tilde{J}\{n_{2}-n_{1}\exp \left[ \hbar \beta \left( \omega \left(
t\right) +p\Delta n-\omega _{21}+\frac{\omega _{st}}{2}\right) \right] \}
\notag \\
&&\pm \frac{n_{2}}{T_{1}}  \label{eq:elpop}
\end{eqnarray}%
where $\beta =1/k_{B}T$, $n_{1}+n_{2}=1$, $\sigma _{a}$ is the cross section
at the maximum of the absorption band, and we added term "$\pm n_{2}/T_{1}$"
taking the lifetime $T_{1}$ of the excited state into account.

In case of fast vibrational relaxation, Eq.(\ref{eq:epsilon(Omega)}) becomes

\begin{align}
\varepsilon (\Omega ,t)& =1+ip_{12}\left( \frac{\varepsilon _{b}+2}{3}%
\right) \sqrt{\frac{\pi }{2\sigma _{2s}}}\times  \notag \\
& \times \sum_{j=1,2}(-1)^{j+1}n_{j}\left( t\right) w[\frac{\Omega -\omega
_{21}+p\Delta n(t)+\delta _{2j}\omega _{st}}{\sqrt{2\sigma _{2s}}}]
\label{eq:epsilon1}
\end{align}

\section{Excitation by chirped pulses compensating "local field" detuning}

\label{sec:chirp}

Eqs. (\ref{eq:rhojjfin}) and even (\ref{eq:elpop}) for populations are
nonlinear equations where the transition frequencies are the functions of
the electronic states populations. So, their solution in general case is not
a simple problem. However, one can use pulses that are suitably chirped
(time-dependent carrier frequency) to compensate for a change of frequency
of the optical transition in time induced by the pulses themselves. This
idea was proposed in studies of a two-state system in relation to Rabi
oscillations in inter-subband transitions in quantum wells [28] and for
obtaining efficient stimulated Raman adiabatic passage (STIRAP) in molecules
in a dense medium [29].

Let us assume that we use suitably chirped pulses compensating the "local
field" detuning $p\Delta n$ that enables us to keep the value of $\omega
_{21}-p\Delta n\left( t\right) -\omega \left( t\right) \equiv \Delta \omega $
as a constant ($\Delta \omega =const$). In that case one can obtain an
integral equation
\begin{eqnarray}
\Delta \left( t\right)  &=&1-\sigma _{a}\left( \frac{\varepsilon _{b}+2}{3}%
\right) ^{2}\int_{0}^{t}dt^{\prime }\frac{\tilde{J}(t^{\prime })\Delta
\left( t^{\prime }\right) }{\sqrt{1-S^{2}\left( t-t^{\prime }\right) }}
\notag \\
&&\times \sum_{j=1}^{2}\exp [-\frac{\left( \Delta \omega -\delta _{j2}\omega
_{st}\right) ^{2}}{2\sigma _{2s}}\frac{1-S\left( t-t^{\prime }\right) }{%
1+S\left( t-t^{\prime }\right) }]  \label{eq:Delta(t)}
\end{eqnarray}%
for the dimensionless non-equilibrium population difference $\Delta \left(
t\right) \equiv \Delta ^{\prime }\left( \Delta \omega ,t\right) /\Delta
^{\prime \left( 0\right) }\left( \Delta \omega \right) $, the effective
methods of the solution of which were developed in Refs. [8,25].

For fast vibrational relaxation, using Eq.(\ref{eq:elpop}), we get

\begin{eqnarray}
\frac{dn_{1,2}}{dt} &=&\pm \sigma _{a}\left( \frac{\varepsilon _{b}+2}{3}%
\right) ^{2}\tilde{J}(t)\exp (-\frac{\Delta \omega ^{2}}{2\sigma _{2s}}%
)\{n_{2}  \notag \\
&&\times \exp [-\frac{\hbar \beta }{2}(\omega _{st}-2\Delta \omega
)]-n_{1}\}\pm \frac{n_{2}}{T_{1}}  \label{eq:elpop2}
\end{eqnarray}

\subsection{Near-zero dielectric function of dense collection of molecules
excited with laser pulse}

\label{sec:laser_pulse}

\begin{figure}[tbp]
\begin{center}
\includegraphics[
height=3.0009in,
width=2.9231in
]{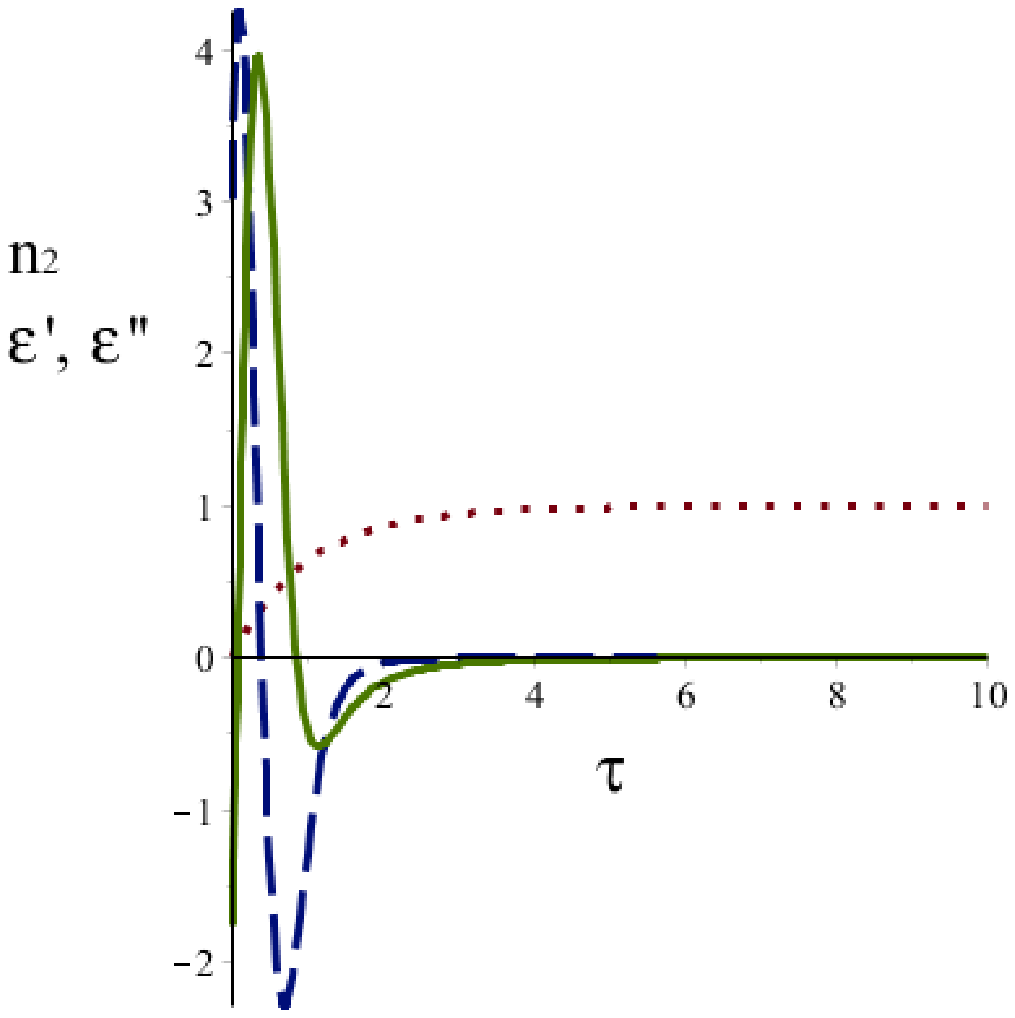}
\end{center}
\caption{Population of the excited state (dots), and real (solid line) and
imaginary (dashed line) parts of the dielectric function as functions of
time.}
\label{fig:n2_eps}
\end{figure}

In this section we shall use Eqs.(\ref{eq:epsilon1}) and (\ref{eq:elpop2})
to demonstrate obtaining near-zero dielectric function in non-steady-state
regime. We shall consider a dense collection of molecules ($N\sim
10^{21}cm^{-3}$ [6]) with parameters close to those of molecule LD690 [8]: $%
\sqrt{\sigma _{2s}}=546cm^{-1}$, $D_{12}\sim 10^{-17}$ CGSE that gives $%
\omega _{st}=1420cm^{-1}$,\ $p_{12}=\allowbreak 2107.\,\allowbreak 2cm^{-1}$%
. We shall put $\varepsilon _{b}=1$ [6] and $\Delta \omega =-420cm^{-1}$.
Fig.\ref{fig:n2_eps} shows the population of excited electronic state $n_{2}$
and the real $\varepsilon ^{\prime }(\Omega ,t)$ and imaginary $\varepsilon
^{\prime \prime }(\Omega ,t)$ parts of $\varepsilon (\Omega ,t)$ for $\Omega
-\omega _{21}=-2.\,\allowbreak 040\,5\sqrt{2\sigma _{2s}}$ during the action
of a rectangular light pulse of power density $\tilde{J}$ that begins at $t=0
$.
Here we denoted
\begin{equation}
W=\sigma _{a}\left( \frac{\varepsilon _{b}+2}{3}\right) ^{2}\exp (-\frac{%
\Delta \omega ^{2}}{2\sigma _{2s}})\tilde{J}  \label{eq:J(t)}
\end{equation}%
- the probability of the optical transitions induced by external field, and $%
\tau =Wt$ - dimensionless time. We put $WT_{1}=1000$. Fig.\ref{fig:n2_eps}
illustrates non-steady-state near-zero dielectric permittivity. As
population $n_{2}$ approaches to $1$, dielectric permittivity approaches to
zero.

\section{Application to exciton compensation of Coulomb blocking (ECCB) in
conduction nanojunctions}

\label{sec:sections}

In Ref. [9] we studied the influence of both exciton effects and Coulomb
repulsion on current in nanojunctions. We showed that dipolar
energy-transfer interactions between the sites in the wire can at high
voltage compensate Coulomb blocking for particular relationships between
their values. Although in free exciton systems dipolar interactions $J$ ($%
\lesssim 0.01-0.1eV$ [30]) are considerably smaller than on-site Coulomb
interaction $U$ (characteristically $U\sim 1$ eV [31]) the former may still
have strong effects under some circumstances, e.g. in the vicinity of
metallic structures in or near the nanojunctions. In such cases dipolar
interactions may be enhanced. The enhancement of the dipole-dipole
interaction calculated using finite-difference time-domain simulation for
the dimer of silver spheres, and within the quasistatic approximation for a
single sphere, reached the value of $0.13$ eV for nanosphere-shaped metallic
contacts [9] that was smaller than $U$. In addition, this enhancement was
accompanied by metal induced damping of excitation energy.

In this section we show that purely organic materials characterized by low
losses with near-zero dielectric permittivities will enable us easily to
obtain $J\sim 1$ eV$\sim U$. We shall consider a nanojunction consisting of
a two site quantum dot wire between two metal leads with applied voltage
bias. The junction is found into organic material with dielectric
permittivity $\varepsilon $. The quantum dots of the wire posses dipole
moments $\mathbf{D}_{1}$ and $\mathbf{D}_{2}$. The point dipoles are
positioned at points $\mathbf{r}_{1}$ and $\mathbf{r}_{2}$, respectively,
and oscillate with frequency $\Omega $. The interaction energy between
dipoles $1$ and $2$ can be written in a symmetrized form as $J_{12}+J_{21}$
where
\begin{align}
J_{12}& =-\frac{1}{2}\mathbf{D}_{1}\cdot \mathbf{E}_{2}(\mathbf{r}%
_{1},\Omega ,t)  \label{eq:U_1D} \\
J_{21}& =-\frac{1}{2}\mathbf{D}_{2}\cdot \mathbf{E}_{1}(\mathbf{r}%
_{2},\Omega ,t)  \label{eq:U_D1}
\end{align}%
$\mathbf{E}_{2}(\mathbf{r}_{1},\Omega ,t)$ $\sim \mathbf{D}_{2}$ is the
electric field at a point $\mathbf{r}_{1}$ induced by the dipole $\mathbf{D}%
_{2}$, etc.

The electric field is given by Coulomb's law%
\begin{equation}
\mathbf{E}(\mathbf{r},\Omega ,t)=\frac{1}{\varepsilon (\Omega ,t)}\int \rho
_{i}(\mathbf{r}^{\prime })\frac{\mathbf{r-r}^{\prime }}{\left\vert \mathbf{%
r-r}^{\prime }\right\vert ^{3}}d^{3}\mathbf{r}^{\prime }  \label{eq:Evar}
\end{equation}%
that corresponds to the electrostatic approximation. Such extension of the
electrostatic formula is possible due slow changes in time of $\varepsilon
(\Omega ,t)$ (see above). Here the external charge density $\rho _{i}(%
\mathbf{r}^{\prime })$ due to the presence of dipole $\mathbf{D}_{i}$ can be
written as [32] $\rho _{i}(\mathbf{r}^{\prime })=-\mathbf{D}_{i}\cdot
\mathbf{\nabla }_{\mathbf{r}^{\prime }}\delta (\mathbf{r}^{\prime }-\mathbf{r%
}_{i})$ (we consider a point dipole positioned at point $\mathbf{r}_{i}$).
One can show that $U_{12}=U_{21}\equiv \frac{1}{2}\hbar J(\Omega ,t)$. This
can be expected from the reciprocity theorem [33], according to which the
fields of two dipoles $\mathbf{D}_{1}$ and $\mathbf{D}_{2}$ at positions $%
\mathbf{r}_{1}$ and $\mathbf{r}_{2}$ and oscillating with the same frequency
$\Omega $ are related as $\mathbf{D}_{1}\cdot \mathbf{E}_{2}(\mathbf{r}%
_{1},\Omega )=\mathbf{D}_{2}\cdot \mathbf{E}_{1}(\mathbf{r}_{2},\Omega )$.
If the dipoles are oriented parallel to the symmetry axis of the junction
\cite{Shishodia11}, the dipole-dipole interaction is given by $J(\Omega
,t)=J_{vac}/\varepsilon (\Omega ,t)$ where $J_{vac}=-2D_{1}D_{2}\left\vert
\mathbf{r}_{1}\mathbf{-r}_{2}\right\vert ^{-3}$ is the dipole-dipole
interaction in vacuum. The bottom of Fig.\ref{fig:ECCBswitch} shows $J$ as a
function of time for a medium with dielectric function given by Fig.\ref%
{fig:n2_eps}. Putting $D_{1}=D_{2}=25D$ and $\left\vert \mathbf{r}_{1}%
\mathbf{-r}_{2}\right\vert =5nm$, one gets $\left\vert J_{vac}\right\vert
=0.006\,25eV$, and the value of $\left\vert J(\Omega ,t)\right\vert
=1.\,\allowbreak 660\,2eV$ for $\tau =10$.

\subsection{Calculation of current. Optical switches based on ECCB.}

\begin{figure}[tbp]
\begin{center}
\includegraphics[
height=2.4803in,
width=3.3883in
]{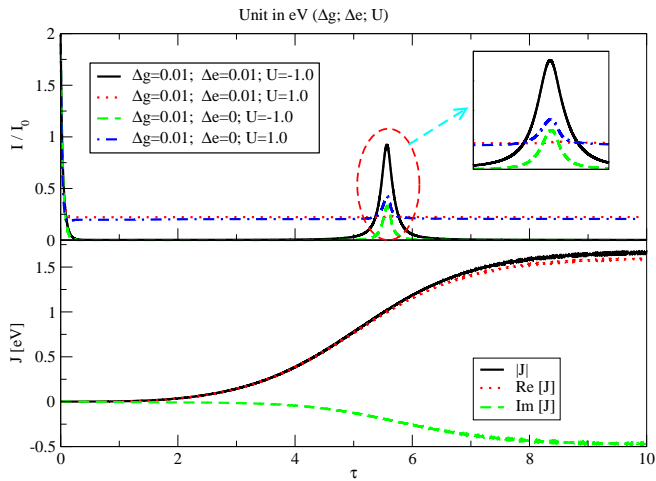}
\end{center}
\caption{(Color online) Laser induced current $I/I_{0}$ (top) and the
dipole-dipole interaction in an organic material $J[eV]$ (bottom) displayed
as functions of $\protect\tau $. Other parameters are identical to those of
Fig.\protect\ref{fig:n2_eps}.}
\label{fig:ECCBswitch}
\end{figure}

Let us calculate current through the two site quantum dot nanojunction
described in the beginning of this section using approach of Ref. [9] where
the dipole-dipole interaction between quantum dots of the wire is defined by
$J(\Omega ,t)=J_{vac}/\varepsilon (\Omega ,t)$ (see above). The Hamiltonian
of the wire, Eq.(3) of Ref. [9], contained both the energy
\begin{equation}
H_{exc-exc}=\hbar J(\Omega ,t)b_{1}^{\dag }b_{2}+H.c.  \label{eq:H_HL}
\end{equation}%
and electron transfer interactions written in the resonance approximation
\begin{equation}
H_{el-el}=-\sum_{f=g,e}\Delta _{f}(\hat{c}_{2f}^{\dag }\hat{c}_{1f}+\hat{c}%
_{1f}^{\dag }\hat{c}_{2f})  \label{eq:el_transfer}
\end{equation}%
The operators $b_{m}^{\dag }={c}_{me}^{\dag }{c}_{mg}$ and $b_{m}={c}%
_{mg}^{\dag }{c}_{me}$ are exciton creation and annihilation operators on
the molecular sites $m=1,2$. The Hamiltonian of the Coulomb interactions is
expressed as
\begin{equation*}
H_{cou}=\dfrac{U}{2}\sum_{m=1,2}N_{m}(N_{m}-1)
\end{equation*}%
with $N_{m}={n}_{mg}+{n}_{me}$. Since in the medium with near-zero
dielectric permittivities both exciton-exciton interaction $J$ and on-site
Coulomb interaction $U$ can achieve the value of about $1$ $eV$ (see above),
we account and add the additional two off-resonance terms to $H_{exc-exc}$
and $H_{el-el}$ respectively, as
\begin{align}
H_{non-exe-exe}& =\hbar J(\Omega ,t)b_{1}^{\dag }b_{2}^{\dag }+H.c.~,
\label{eq:H_non-HL} \\
H_{non-el-el}& =-\sum_{\substack{ f,f^{\prime }=g,e \\ f\neq f^{\prime }}}%
\Delta _{ff^{\prime }}(\hat{c}_{2f}^{\dag }\hat{c}_{1f^{\prime }}+\hat{c}%
_{1f^{\prime }}^{\dag }\hat{c}_{2f})  \label{eq:el_transfer_nonres}
\end{align}%
Eq.~\ref{eq:H_non-HL} is so called non-Heitler-London term [35] taking into
account creation and annihilation for excitation simultaneously at two sites
(quantum dots). In this relation the following question arises: "does the
effect of ECCB survive for such large values of $\hbar J\sim 1$ eV$\sim U$?"
\begin{figure}[tbp]
\begin{center}
\includegraphics[
height=2.5313in,
width=3.5535in
]{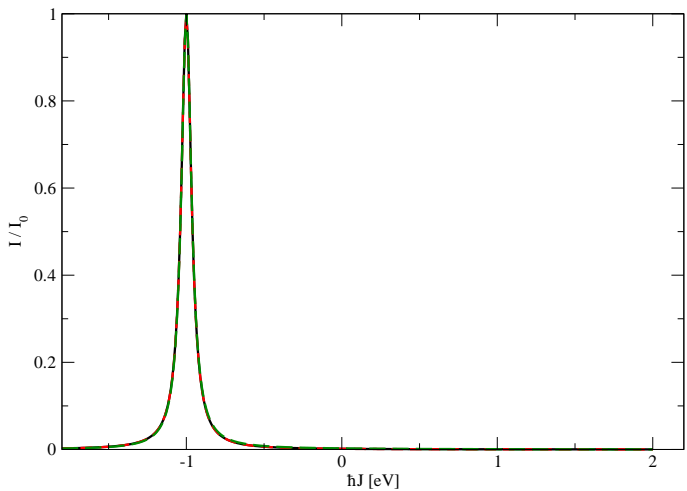}
\end{center}
\caption{(Color online) Current $I/I_{0}$ displayed as function of the
energy-transfer coupling $J$ for $U=1$ $eV$ and $\Delta _{f}=\Delta
_{ff^{\prime }}\equiv \Delta =0.01eV$. The current shows a maximum at $%
U=-\hbar J$. Solid black line - calculations using Eqs.(\protect\ref{eq:H_HL}%
) and (\protect\ref{eq:el_transfer}) for energy and electron transfer,
respectively; red dashed line - calculations using Eqs.(\protect\ref{eq:H_HL}%
) and (\protect\ref{eq:H_non-HL}) for energy, and Eq.(\protect\ref%
{eq:el_transfer}) for electron transfer; green dashed line - calculations
using Eqs.(\protect\ref{eq:H_HL}) and (\protect\ref{eq:H_non-HL}) for
energy, and Eqs.(\protect\ref{eq:el_transfer}) and (\protect\ref%
{eq:el_transfer_nonres}) for electron transfer. Comparison of these lines
displays small effects of non-resonance contributions.}
\label{fig:ECCBlargeJ}
\end{figure}
Fig.\ref{fig:ECCBlargeJ} shows that the ECCB does survive for large values
of $J\sim 1$ eV. We put the bias voltage $V_{bs}=8$ $eV$ and the rate of
charge transfer from a quantum dot to the corresponding lead $\Gamma =0.02eV$
in our simulations, and denoted the unit of current as $I_{0}=\dfrac{e\Gamma
}{\hbar }$ ($e$ is the charge of one electron). Fig.\ref{fig:ECCBswitch}
shows current through the nanojunction during the action of the rectangular
lase pulse with parameters given in Section\ref{sec:laser_pulse} on the host
organic material.

One can see dramatic increasing the current when $\hbar J$ approaches to $-U$
for $\Delta _{g}=\Delta _{e}$, and to $\pm U$ for $\Delta _{g}\neq 0$ and $%
\Delta _{e}=0$. After this moment the current decreases in spite of
increasing $J$, since its value exceeds that of $U$. So, current exists
during the time that is much shorter than the pulse duration. As a matter of
fact, Fig.\ref{fig:ECCBswitch} illustrates a new type of optical switches
based on the effect of the exciton compensation of Coulomb blocking - ECCB
switches.

\section{Bistability}

\begin{figure}[tbp]
\begin{center}
\includegraphics[
height=3.6902in,
width=2.2546in
]{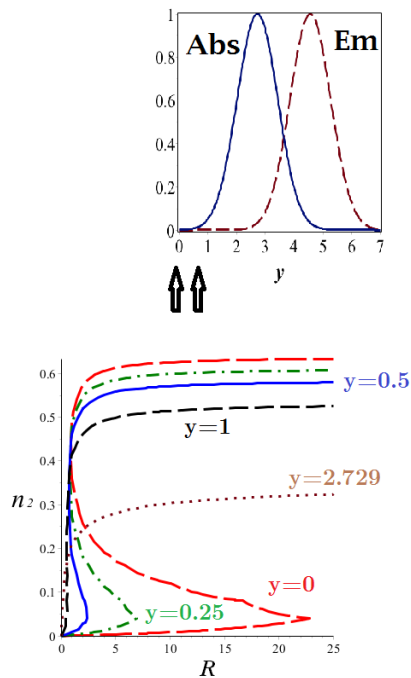}
\end{center}
\caption{(Color online) Dependence of excited state population $n_{2}$ on
power density of the exciting radiation $\tilde{J}$ at different detunings $%
\protect\omega _{21}-\protect\omega $. Dimensionless parameters are $R=%
\protect\sigma _{a}\tilde{J}T_{1}$ and $y=(\protect\omega _{21}-\protect%
\omega )/\protect\sqrt{2\protect\sigma _{2s}}$. Parameters $\protect\sqrt{%
\protect\sigma _{2s}}$, $\protect\omega _{st}$, $p_{12}$ and $\protect%
\varepsilon _{b}$ are identical to those of Section\protect\ref%
{sec:laser_pulse}. Inset: Equilibrium spectra of the absorption (Abs) and
the emission (Em); the arrows limit the frequency interval where calculated
excited state populations $n_{2}$ show bistability.}
\label{fig:n2_bistability}
\end{figure}

If one does not use suitably chirped pulses that compensate for a change of
frequency of the optical transition in time induced by the pulses themselves
(see Section\ref{sec:chirp}), Eqs. (\ref{eq:rhojjfin}) and (\ref{eq:elpop})
for populations become nonlinear equations and can demonstrate a bistable
behavior. Fig.\ref{fig:n2_bistability} shows steady-state solutions of Eq.(%
\ref{eq:elpop}) for $n_{2}$ as a function of the power density of the
exciting radiation $\tilde{J}$ at different detunings $\omega _{21}-\omega $%
.
One can see that each value of $\tilde{J}$ within the corresponding interval
produces three different solutions of Eq.(\ref{eq:elpop}) for dimensionless
detunings $y=(\omega _{21}-\omega )/\sqrt{2\sigma _{2s}}=0,$ $0.25$ and $0.5$%
, however, only lower and upper branches are stable [36]. Such detunings
correspond to the excitation at the short-wave part of the equilibrium
absorption spectrum (see the Inset to Fig.\ref{fig:n2_bistability}). As the
excited state population increases, the spectrum exhibits the blue shift
(see Eq.(\ref{eq:elpop}) that should essentially contribute to the
absorption. As a matter of fact, the bistable behavior of the population
arises from the dependence of the resonance frequency of the molecules in
dense medium on the number of excited molecules. In contrast, larger $y=1$, $%
2.729$ correspond to the excitation closer to the central part of the
equilibrium absrption spectrum. In that case the blue shift produces lesser
increasing the absorption and even can decrease it (for $y=2.729$), so that
the bistable behavior disappears.

Furthermore, the excitation of surface plasmon polaritons at the organic
thin film/air interface is possible for substantially strong negative values
of dielectric function $\varepsilon (\Omega )$ [6,37].The bistable behavior
of the population results in the bistable behavior of $\varepsilon (\Omega )$
(see Eqs.(\ref{eq:epsilon(Omega)}) and (\ref{eq:epsilon1})), and as a
consequence, a bistable behavior of the dispersion relations for surface
plasmon polaritons at the organic thin film/air interface under the laser
irradiation. This issue will be considered elsewhere.

\section{Conclusion}

In this work we have developed a theory of nonlinear non-steady-state
organic plasmonics with strong laser pulses. We have demonstrated the
bistabile response of the electron-vibrational model of organic materials in
condensed phase that leads to the bistability of their plasmonic properties.
Specifically, bistability in the regime of the surface plasmon polariton
propagation at the organic thin film/air interface may be used for new types
of optical switches. We have proposed to use non-steady-state organic
plasmonics for the enhancement of intersite dipolar energy-transfer
interaction in the quantum dot wire that influences on electron transport
through nanojunctions. Such interactions can compensate Coulomb repulsions
for particular conditions. We propose the exciton control of Coulomb
blocking in the quantum dot wire based on the non-steady-state near-zero
dielectric permittivity of the organic host medium, and a new type of
optical switches - ECCB switches. Our current calculations were carried out
for a value of $J(\Omega ,t)$ corresponding to fixed frequency $\Omega $ $%
(=\omega _{21}-2.\,\allowbreak 040\,5\sqrt{2\sigma _{2s}})$. The extension
of the calculations of current to frequency dependent $J(\Omega ,t)$ will be
made elsewhere.

\acknowledgments     

We gratefully acknowledge support by the US-Israel Binational Science
Foundation (BF, grant No. 2008282).

\section{Appendix}

Let us generalize equations of Ref. [8] to the dipole-dipole intermolecular
interactions in the condensed matter. The latter are described by
Hamiltonian [18,35] $\hat{H}_{int}=\hbar \sum_{m\neq n}J_{mn}b_{m}^{\dag
}b_{n}+H.c.$ (compare with Eq.(\ref{eq:H_HL})). Using the Heisenberg
equations of motion, one obtains that $\hat{H}_{int}$ gives the following
contribution to the change of the expectation value of excitonic operator $%
b_{k}$ in time%
\begin{eqnarray}
\frac{d}{dt}\langle b_{k}\rangle  &\sim &\frac{i}{\hbar }\langle \lbrack
\hat{H}_{int},b_{k}]\rangle \equiv \frac{i}{\hbar }Tr([\hat{H}%
_{int},b_{k}]\rho )  \notag \\
&=&i2\sum\limits_{n\neq k}J_{kn}\langle (\hat{n}_{k2}-\hat{n}%
_{k1})b_{n}\rangle   \label{eq:Heis1}
\end{eqnarray}%
where $\rho $ is the density matrix, $\left\vert k1\right\rangle \equiv
\left\vert kg\right\rangle ,$ $\left\vert k2\right\rangle \equiv \left\vert
ke\right\rangle $, $\hat{n}_{k1}=b_{k}b_{k}^{\dag }$, and $\hat{n}%
_{k2}=b_{k}^{\dag }b_{k}$ is the exciton population operator. Considering an
assembly of identical molecules, one can write $\langle b_{k}\rangle =\rho
_{21}\left( \alpha ,t\right) $ [29] if averaging in Eq.(\ref{eq:Heis1}) is
carried out using density matrix $\rho \left( \alpha ,t\right) $. Consider
the expectation value $\langle (\hat{n}_{k2}-\hat{n}_{k1})b_{n}\rangle =Tr[(%
\hat{n}_{k2}-\hat{n}_{k1})b_{n}\rho \left( \alpha ,t\right) ]$ for $n\neq k.$
Due to fast dephasing (see Section\ref{sec:model}), it makes sense to
neglect all correlations among different molecules [18], and set $\langle (%
\hat{n}_{k2}-\hat{n}_{k1})b_{n}\rangle =\langle \hat{n}_{k2}-\hat{n}%
_{k1}\rangle \langle b_{n}\rangle $. Here from dimension consideration one
expectation value should be calculated using density matrix $\rho \left(
\alpha ,t\right) $, and another one - using $\langle \rho \rangle \left(
t\right) =\int \rho \left( \alpha ,t\right) d\alpha $. Bearing in mind fast
dephasing, we choose option $\langle \hat{n}_{k2}-\hat{n}_{k1}\rangle
\langle b_{n}\rangle =(n_{2}-n_{1})\rho _{21}\left( \alpha ,t\right) $ that
gives the most contribution and results in the agreement with experimental
spectra of molecular thin films. Another option $\langle \hat{n}_{k2}-\hat{n}%
_{k1}\rangle \langle b_{n}\rangle =-\Delta ^{\prime }\left( \alpha ,t\right)
\langle \rho \rangle _{21}\left( t\right) $ is more suitable to the creation
of delocalized collective states, and similar to the procedure used for the
derivation of the semiconductors Bloch equations [38,39].

It remains to calculate $2\sum_{n\neq k}J_{kn}=2\lim_{\mathbf{k}\rightarrow
0}J(\mathbf{k})$ on the right-hand side of Eq.(\ref{eq:Heis1}) that is
conveniently calculated in $\mathbf{k}$ space where $J(\mathbf{k}%
)=\sum_{n\neq k}J_{kn}\exp (-i\mathbf{k}\cdot \mathbf{r}_{n})$, $\mathbf{r}%
_{n}$ denotes the position of the $n$th molecule. Bearing in mind that $%
\hbar J_{kn}\equiv \frac{1}{2}\mathbf{D}_{k}\cdot T_{kn}\cdot \mathbf{D}_{n}$
where $T_{kn}$ is the dipole-dipole tensor, and using $T(\mathbf{k})=-4\pi
N/3$, Eq.(16.20b) of Ref. [18] for a transverse field (see also [40]), we
get $2\sum_{n\neq k}J_{kn}=-p$. This yields $\partial \rho _{21}\left(
\alpha ,t\right) /\partial t\sim ip\Delta n\rho _{21}\left( \alpha ,t\right)
$. Adding term "$ip\Delta n\rho _{21}\left( \alpha ,t\right) $" to the
right-hand side of Eq.(9) of Ref.[8], and using the procedure described
there, we get Eq.(\ref{eq:rhojjfin}).

\textbf{References}%

[1] J.~B. Khurgin, ``How to deal with the loss in plasmonics and
metamaterials,''  \emph{Nature Nanotechnology}, vol.~10, pp. 2--6, 2015.%

[2] A.~J. Hoffman, L.~Alexeev, S.~S. Howard, K.~J. Franz, D.~Wasserman,
V.~A.  Podolskiy, E.~E. Narimanov, D.~L. Sivco, and C.~Gmachl, ``Negative
refraction  in semiconductor metamaterials,'' \emph{Nature Materials},
vol.~6, pp.  946--950, 2007.

[3] F.~H.~L. Koppens, D.~E. Chang, and F.~J.~G. de~Abajo, ``Graphene
plasmonics: A  platform for strong light-matter interaction,'' \emph{Nano
Letters}, vol.~11,  pp. 3370--3377, 2011.

[4] J.~Chen, M.~Badioli, P.~Alonso-Gonzalez, S.~Thongrattanasiri, F.~Huth,
J.~Osmond, M.~Spasenovic, A.~Centeno, A.~Pesquera, P.~Godignon, A.~Z.
Elorza,  N.~Camara, F.~J.~G. de~Abajo, R.~Hillenbrand, and F.~H.~L. Koppens,
``Optical  nano-imaging of gate-tunable graphene plasmons,'' \emph{Nature},
vol. 487,  pp. 77--81, 2012.

[5] Z.~Fei, A.~S. Rodin, G.~O. Andreev, W.~Bao, A.~S. McLeod, M.~Wagner,
L.~M.  Zhang, Z.~Zhao, M.~Thiemens, G.~Dominguez, M.~M. Fogler, A.~H.~C.
Neto, C.~N.  Lau, F.~Keilmann, and D.~N. Basov, ``Gate-tuning of graphene
plasmons  revealed by infrared nano-imagine,'' \emph{Nature}, vol. 487, pp.
82--85,  2012.

[6] L.~Gu, J.~Livenery, G.~Zhu, E.~E. Narimanov, and M.~A. Noginov, ``Quest
for  organic plasmonics,'' \emph{Applied Phys. Lett.}, vol. 103, p. 021104,
2013.

[7] T.~U. Tumkur, J.~K. Kitur, L.~Gu, G.~Zhu, and M.~A. Noginov, in \emph{%
Abstracts of NANOMETA 2013}, Seefeld, Austria, 2013, p. FRI3o.6.

[8] B.~D. Fainberg, ``Nonperturbative analytic approach to interaction of
intense  ultrashort chirped pulses with molecules in solution: Picture of
''moving''  potentials,'' \emph{J. Chem. Phys.}, vol. 109, no.~11, pp.
4523--4532, 1998.

[9] G.~Li, M.~S. Shishodia, B.~D. Fainberg, B.~Apter, M.~Oren, A.~Nitzan,
and  M.~Ratner, ``Compensation of coulomb blocking and energy transfer in
the  current voltage characteristic of molecular conduction junctions,''
\emph{Nano Letters}, vol.~12, pp. 2228--2232, 2012.

[10] B.~D. Fainberg and B.~S. Neporent, \emph{Opt. Spectrosc.}, vol.~48, p.
393,  1980, [Opt. Spektrosk., v. 48, 712 (1980)].

[11] B.~D. Fainberg and I.~B. Neporent, \emph{Opt. Spectrosc.}, vol.~61,
p.~31,  1986, [Opt. Spektrosk., v. 61, 48 (1986)].

[12] B.~D. Fainberg and I.~N. Myakisheva, \emph{Sov. J. Quant. Electron.},
vol.~17,  p. 1595, 1987, [Kvantovaya Elektron. (Moscow), v. 14, 2509 (1987)].

[13] ------, \emph{Opt. Spectrosc.}, vol.~66, p. 591, 1989, [Opt.
Spektrosk., v. 66,  1012 (1989)].

[14] B.~D. Fainberg, \emph{Opt. Spectrosc.}, vol.~60, p.~74, 1986, [Opt.
Spektrosk.,  v. 60, 120 (1986)].

[15] R.~F. Loring, Y.~J. Yan, and S.~Mukamel, \emph{J. Chem. Phys.},
vol.~87, p.  5840, 1987.

[16] W.~Vogel, D.-G. Welsh, and B.~Wilhelmi, \emph{Phys. Rev. A}, vol.~37,
p. 3825,  1988.

[17] B.~D. Fainberg, \emph{Opt. Spectrosc.}, vol.~65, p. 722, 1988, [Opt.
Spektrosk., vol. 65, 1223, 1988].

[18] S.~Mukamel, \emph{Principles of Nonlinear Optical Spectroscopy}.\hskip %
1em plus  0.5em minus 0.4em\relax New York: Oxford University Press, 1995.

[19] V.~V. Khizhnyakov, \emph{Izv. Akad. Nauk SSSR, Ser. Fiz.}, vol.~52, p.
765,  1988.

[20] B.~D. Fainberg, \emph{Opt. Spectrosc.}, vol.~58, p. 323, 1985, [Opt.
Spektrosk.  v. 58, 533 (1985)].

[21] Y.~J. Yan and S.~Mukamel, \emph{Phys. Rev. A}, vol.~41, p. 6485, 1990.

[22] B.~D. Fainberg, \emph{Opt. Spectrosc.}, vol.~68, p. 305, 1990, [Opt.
Spektrosk., vol. 68, 525, 1990].

[23] B.~Fainberg, \emph{Phys. Rev. A}, vol.~48, p. 849, 1993.

[24] B.~D. Fainberg, \emph{Opt. Spectrosc.}, vol.~67, p. 137, 1989, [Opt.
Spektrosk., v. 67, 241 (1989)].

[25] ------, ``Non-linear polarization and spectroscopy of vibronic
transitions in  the field of intensive ultrashort pulses,'' \emph{Chem. Phys.%
}, vol. 148, pp.  33--45, 1990.

[26] M.~E. Crenshow, M.~Scalora, and C.~M. Bowden, ``Ultrafast intrinsic
optical  switching in dense medium of two-level atoms,'' \emph{Phys. Rev.
Lett.},  vol.~68, pp. 911--914, 1992.

[27] M.~Abramowitz and I.~Stegun, \emph{Handbook on Mathematical Functions}.%
\hskip
1em plus 0.5em minus 0.4em\relax New York: Dover, 1964.

[28] A.~A. Batista and D.~S. Citrin, ``Rabi flopping in a two-level system
with a  time-dependent energy renormalization: Intersubband transitions in
quantum  wells,'' \emph{Phys. Rev. Lett.}, vol.~92, no.~12, p. 127404, 2004.

[29] B.~D. Fainberg and B.~Levinsky, ``Stimulated raman adiabatic passage in
a dense  medium,'' \emph{Adv. Phys. Chem.}, vol. 2010, p. 798419, 2010.

[30] S.~Mukamel and D.~Abramavicius, ``Many-body approaches for simulating
coherent  nonlinear spectroscopies of electronic and vibrational excitons,''
\emph{Chem. Rev.}, vol. 104, pp. 2073--2098, 2004.

[31] H.~Thomann, L.~R. Dalton, M.~Grabowski, and T.~C. Clarke, ``Direct
observation  of coulomb correlation effects in polyacetylene,'' \emph{Phys.
Rev. B},  vol.~31, no.~5, pp. 3141--3143, 1985.

[32] X.~M. Hua, J.~I. Gersten, and A.~Nitzan, ``Theory of energy transfer
between  molecules near solid particles,'' \emph{J. Chem. Phys.}, vol.~83,
no.~7, pp.  3650--3659, 1985.

[33] L.~D. Landau and E.~M. Lifshitz, \emph{Electrodynamics of Continuous
Media}.\hskip 1em plus 0.5em minus 0.4em\relax New York: Pergamon Press,
1960.

[34] M.~S. Shishodia, B.~D. Fainberg, and A.~Nitzan, ``Theory of energy
transfer  interactions near sphere and nanoshell based plasmonic
nanostructures,'' in  \emph{Plasmonics: Metallic Nanostructures and Their
Optical Properties IX. Proc. of SPIE}, M.~I. Stockman, Ed.\hskip 1em plus
0.5em minus 0.4em\relax
Bellingham, WA: SPIE, 2011, vol. 8096, p. 8096 1G.

[35] A.~S. Davydov, \emph{Theory of Molecular Excitons}.\hskip 1em plus
0.5em minus  0.4em\relax New York: Plenum, 1971.

[36] N.~N. Bogoliubov and Y.~A. Mitropolskyi, \emph{Asymptotic methods in
the theory of non-linear oscillations}.\hskip 1em plus 0.5em minus 0.4em%
\relax New York:  Gordon and Breach, 1961.

[37] H.~Raether, \emph{Surface Plasmons on Smooth and Rough Surfaces and on
Gratings}.\hskip 1em plus 0.5em minus 0.4em\relax Berlin: Springer-Verlag,
1986.

[38] M.~Lindberg and S.~W.Koch, ``Effective bloch equations for
semiconductors,''  \emph{Phys. Rev. B}, vol.~38, no.~5, pp. 3342--3350, 1988.

[39] H.~Haug and S.~W. Koch, \emph{Quantum theory of the optical and
electronic properties of semiconductors}.\hskip 1em plus 0.5em minus 0.4em%
\relax
Singapore: World Scientific, 2001.

[40] A.~Gonzalez-Tudela, D.~Martin-Cano, E.~Moreno, L.~Martin-Moreno,
C.~Tejedor,  and F.~J. Garcia-Vidal, ``Dipolar sums in the primitive cubic
lattices,''  \emph{Phys.Rev.}, vol.~99, no.~4, pp. 1128--1134, 1955.


\end{document}